\documentclass[epj
]{svjour}
\usepackage{graphicx}
\begin{document}
\titlerunning{{Reply to ``What Faster-than-traffic Characteristic Speeds Mean...''}}
\title{Reply to ``What Faster-than-traffic Characteristic Speeds Mean for Vehicular Traffic Flow''}
\author{Dirk Helbing
}                     
%
%
\institute{ETH Zurich, UNO D11, Universit\"atstr. 41, 8092 Zurich, Switzerland}
\date{Received: date / Revised version: date}
%
\abstract{This contribution replies to H. M. Zhang's paper  ``What Faster-than-traffic Characteristic Speeds Mean for Vehicular Traffic Flow'', which comments on the manuscript ``On the Controversy around Daganzo's Requiem for and Aw-Rascle's Resurrection of Second-Order Traffic Flow Models'' 
published in this issue of Eurpean Physical Journal B. The author clarifies several points and agrees that suitable experiments can be a way of deciding the controversy on the existence of propagation speeds faster than the speed of vehicles. Moreover, he proposes to aim at an integrated theory, which is both, theoretically consistent and practically relevant.
} 
\maketitle
As vehicle traffic is influenced by decision-making and psychological effects, nobody would expect that traffic models could reach an accuracy comparable to models in physics. Nevertheless, they can help to understand non-trivial properties of traffic flows, to predict and optimize them. In many cases, models do not even {\it aim} at giving an accurate picture of reality. The strength of models lies in the simplification of reality. Approximations are often unavoidable to make theoretical progress,
and from the point of view of engineering and empirical facts, a good approximation can sometimes be preferable to a theoretically consistent model that does not match the data. 
Therefore, theoretical consistency alone is not a good selection criterium for models. 
Regarding traffic flows, for example, there is a long list of empirical observations, which theories should be able to match \cite{TranSci}. This particularly concerns the evolution and spatio-temporal characteristics of congestion patterns. However, there is no doubt that a theoretically consistent model is preferable to a non-consistent one, if both models reproduce empirical facts equally well, considering the number of parameters.
\par
The ``Requiem for second-order fluid approximations of traffic flow''  basically argues along these lines \cite{Daganzo}. In this paper, Daganzo demonstrates that the Payne-Whitham and some other macroscopic traffic models predict behaviors, which are not theoretically consistent and {\it also} not consistent with empirical observations. For example, under certain conditions, vehicles are predicted to move backwards, and congestion fronts tend to smear out much more than they should. These properties result from the finite-order Taylor expansion made, when deriving the model from ``microscopic'' car-following models \cite{Daganzo}. In fact, it has been shown in Ref. \cite{EPJB} that such a gradient expansion is not justified for non-homogeneous traffic flows, and that the appropriate form of macroscopic traffic equations is non-local. Considering this, the previously mentioned inconsistencies disappear. One could hope that 
the theoretical prediction of a propagation speed (``characteristic'') that is faster than the vehicle speed, would also disappear, when the above mentioned Taylor approximation is avoided. However, in contrast to the inconsistency of backward propagating cars, the faster-than-traffic characteristic speed does not only occur for {\it large} changes of the density along the freeway, but also for arbitrarily small deviations from the stationary and homogeneous solution, as a linear instability analysis reveals \cite{EPJB2}. The same results are found for the optimal velocity model, which is a simple car-following model with non-negative velocities, if model parameters and initial conditions are reasonably chosen. That does, of course, not mean that the optimal velocity model and the Payne-Whitham model would behave identically for {\it finite} perturbations of the stationary and homogeneous solution. 
\par
There have been many attempts to understand and overcome the problem of a faster-than-traffic characteristic speed. A good review of them is given by Zhang \cite{Zhang}. In particular, new traffic models were proposed, which do not have 
any faster-than-traffic characteristic speed. However, a satisfactory interpretation {\it why} certain traffic models show a propagation speed faster than the vehicle speed has been lacking. Based on an interpretation of the formulas resulting from a linear stability analysis, Ref.  \cite{EPJB2} now argues that it is not a matter of vehicle {\it interactions}, but of vehicle {\it acceleration}, which causes the faster-than-traffic characteristic speed in the Payne-Whitham model. 
It is shown that, for certain initial conditions in which cars are driving more slowly than the distance-dependent equilibrium speed, the acceleration resulting from the relaxation term may cause cars in the back to accelerate more strongly than the cars in front. As a consequence, it appears that cars in the back accelerate {\it earlier} than the cars in front. However, since this is not an interaction effect, it does not violate the principle of causality. 
\par
Therefore, according to Ref. \cite{EPJB2}, faster-than-traffic characteristic speeds do not constitute a theoretical inconsistency, but just a problem of interpretation.
In fact, in the limit $\tau\rightarrow \infty$, in which the relaxation term vanishes, the characteristic speed
is restricted to the average vehicle speed $V_{\rm e}$ plus the square root of the the derivative $\partial {\cal P}_1/\partial \rho$ of the traffic pressure ${\cal P}_1(\rho)=\rho \theta(\rho)$ with respect to the density $\rho$ (given there is no speed-dependent pressure contribution ${\cal P}_2$). This exactly corresponds to the empirically well supported formula for the speed of sound propagation in gases. As vehicle speeds are not identical when driver-vehicle units are heterogeneous or traffic flows are spatially non-homogeneous, the resulting finite velocity variance $\theta$ implies that a realistic traffic theory should actually {\it have} a characteristic speed faster than the average vehicle speed.
\par
Why should we care about these issues? Because one would like to have a theoretically consistent traffic theory that is capable of explaining as many stylized facts of traffic flows as possible. A first attempt to construct such a comprehensive theory has recently been made \cite{theory}. Although some other car-following models (considering a dependence of a driver's acceleration behavior on vehicle speeds) are known to be more realistic, this theory is based on the optimal velocity model for the sake of analytical tractability. If it turns out that the optimal velocity model cannot adequately represent the most essential features of traffic flows, an equally far-reaching theory would have to be derived from a better car-following model, covering subjects like the micro-macro-link, instabilities  of traffic flows and propagation speeds of perturbations, metastable traffic flows and critical perturbation amplitudes, the observed kinds of spatio-temporal traffic patterns and their conditions of occurence, effects of stochasticity, effects of heterogeneous driver-vehicle units and multi-lane traffic, effects of network flows with merges, diverges, and intersections, fundamental relationships for urban traffic, as well as 
aspects of traffic optimization and control. 
The question is, whether such an alternative theory would lead to a substantial improvement in the understanding of traffic dynamics and to statistically significant improvements in the quantitative prediction of empirical measurements without increasing the number of parameters.
\par
The scientific debate about faster-than-traffic characteristic speeds suggests that the choice of the underlying model is a fundamental issue, as it is claimed that certain models would be theoretically inconsistent and unsuitable for traffic modeling. In this connection, it must be decided, whether the phenomenon of faster-than-traffic characteristic speeds really constitutes a theoretical inconsistency and, if yes, whether it is a negligible effect or a serious shortcoming. Due to the stability of the eigenmode that is associated with the faster-than-traffic characteristic speed \cite{EPJB2}, it seems to be a negligible effect. According to Ref. \cite{EPJB2}, it is also not a theoretical shortcoming, but a problem of interpretation. But maybe, not everybody agrees.
\par
The controversy may be finally decided by further mathematical analysis, by computer simulation of plausible car-following models, and/or by experimental investigations. The problem of experiments is that the faster-than-traffic characteristic speed is expected to occur only over a short time period that is proportional to the inverse relaxation time $1/\tau$. This is expected to complicate measurements. However, effects of a faster-than-traffic characteristic speed may also be observable in experimental setups involving on- or off-ramps. Moreover, experiments could be performed with programmed toy vehicles, which implement certain car-following rules and allow one to modify the parameter values. 
\par
A problem of mathematical analysis and computer simulations is the circumstance that, in the Payne-Whitham and the optimal velocity model both, acceleration and interaction effects are included in the relaxation term. Therefore, effects of vehicle acceleration are not clearly separated from the effects of vehicle interactions. However, for many macroscopic traffic models that are derived from gas-kinetic traffic models, the acceleration term and the interaction term can be separated. The acceleration term typically has the form $(v^0 - V)/\tau$, where $v^0$ is the free vehicle speed and $V$ the actual one. The interaction term corresponds to the remaining contribution of the relaxation term, which is more or less proportional to the velocity variance $\theta$. By varying the relaxation time $\tau$ and the velocity variance $\theta$, one can separately study the impact of the two contributions on the characteristic propagation speeds. 
\par
When aiming at better models, it should be considered that theoretical consistency is just {\it one} aspect, and matching empirical data or stylized facts is at least equally important. It is not enough to find partial answers. In the end, one would like to have an integrated, far-reaching theory, which covers a large set of theoretical as well as empirical requirements, and which is both, theoretically consistent and practically relevant. After all, we are trying to describe the same, measurable reality. It would be quite dissatisfactory, if we ended up with different schools of thought. This could neither convince scientists nor anybody else.

\end{document}